# Explaining the discrepancy in the Chandler period and the prediction of the changes in the Earth's rotation dynamics

Ranjan Vepa[1]

School of Engineering and Material Science, Queen Mary, University of London, London, E1 4NS, UK.

**Abstract**

In this paper, in the first instance the attitude dynamics of the Earth is modelled based on physical principles so as to correctly predict the Chandler wobble and its features such as its period. To this end not only the steady state and dynamic gravity gradient torques were included, but also the effects of centrifugal acceleration on the deformable viscoelastic model of the planet. After validating the Chandler wobble response, the paper seeks to go beyond and predict the attitude response of the Earth not just over a few years but over much longer term. This requires modelling of the energy balance effects of the entire planet as a whole and not just the redistribution of mass and inertia but also the changes in the inertial properties. To this end simpler physical model of the energy balance process was included to demonstrate the sustained changes in the length of day response of the Earth over a long term. The over aim of the paper is to bring into focus the key physical processes responsible for the changes to the attitude dynamics of the Earth over a reasonable longer time frame. The overall responses of the attitude dynamics over, the short and medium terms are presented.

**Keywords**

Chandler wobble, attitude dynamics model, thermodynamic energy balance, length of day, gravity gradient torque, Earth's moments of inertia

**Introduction**

While Earth's rotation and revolution significantly influence its climate, it is also true that climate change and global warming have a greater influence on the Earth's rotational speed than the moon has, which in turn has altered the length of the day. Thus, the coupling between the climate and Earth's rotation is mutual and could be nonlinear. The Milankovitch cycles which include the shape of Earth's orbit, known as eccentricity, the angle Earth's axis is tilted with respect to Earth's orbital plane, known as obliquity and the direction Earth's axis of rotation is pointed, known as precession, cannot by themselves explain current global warming events, particularly the steady increase in $CO_2$, the global temperature, Methane and the steady decrease or arctic sea ice. It has been suggested that variations in the Milankovitch cycles, particularly in the obliquity and in the precession due to features such as the Chandler wobble, apart from the sustained increases in $CO_2$, could result in cyclic variations as well as sustained nonlinear increases in temperature due to second order effects, resulting in changes in the length of day and other related consequences. Some studies to identify the role of climate change on length of day variations have been conducted in a NASA-JPL funded study by Kiani Shahvandi et al (Kiani Shahvandi et al. 2024a, 2024b), using a physics informed neural network (PINN).

NASA which has hitherto funded studies on explaining climate change, Wang and Lee (2024), monitors the climate change parameters using a fleet of 20 satellites and a few CubeSats as mentioned by NASA Science Editorial Team, (2020), Velev (2025) and Velev (2024). Notable amongst these are GRACE-FO (Gravity Recovery and Climate Experiment Follow On): Provides global observations of surface mass changes, which helps monitor things like ice sheet and groundwater changes, NISAR (NASA-ISRO Synthetic Aperture Radar): A joint mission with the Indian Space Research Organisation (ISRO) that monitors Earth's changing ecosystems and frozen regions, Terra: A flagship mission in NASA's Earth Observing System that has been providing long-term data on various Earth systems and GPM (Global Precipitation Measurement): A joint mission with the Japan Aerospace Exploration Agency (JAXA) that provides detailed information about precipitation on a global scale.

---

[1] Corresponding author, email: r.vepa@qmul.ac.uk

The redistribution of mass due to climate change effects the gravitational pull of the Earth and the gravitational model (Howarth, 2025). This is due to shifts in the masses near the poles and resulting polar motion manifests itself as a planetary wobble. The original GRACE mission recorded the Earth's gravitational field and its changes over time in more than 160 months. The follow-on Grace mission, Grace-FO, (Flechtner, 2025), in collaboration with the German, GFZ, (Germany's national centre for Earth system research) was aimed at to obtain precise global and high-resolution models for both the static and the time variable components of the Earth's gravity field. The satellites were in near Polar orbits at an altitude of 485 km.

The aim of this paper is primarily to provide a roadmap that will lead to a high fidelity model for analysing the rotational dynamics of the planet Earth, over the short, medium and longer terms.

**Earth's Rotation Rate**

One of major consequences of the melting of the polar ice caps near the North & South poles, which results in the solid ice turning into liquid (water), followed by both an outward radial as well as an azimuthal shift of the water due to centrifugal and Coriolis effects, is the increase in the Earth's polar MI. The outward movement of the liquid particles is due to the reduction in the inter-particulate forces binding them together in the liquid phase. The consequent increase in the polar MI can be computed from the changes in the zonal harmonics of the estimated gravity model. The increase in the Earth's polar MI, no matter how small results, in a small slowing down of the Earth's rotation rate about its polar axis, changing the overall length of the day as well as the time of exposure to the Sun, both of which increase. This leads to an increase in the temperatures, speeds up the melting of the ice caps and slows the Earth's rotation rate even more. In short, an unstable process! Yet, the increase in the Earth's polar MI, is generally considered to be **long term effect** due to the fact the changes to the deformation of the equatorial bulge takes several hundred years. In **the medium term**, (20-100 years) the MI changes to the Earth due to sea level rise may be computed by an approximate method outlined in Appendix A. In this preliminary model, the influence of the Moon (Tidal effect), the centrifugal forces and the atmospheric effects were not assumed to change and will be included at a later stage. The centrifugal forces disproportionately affect the polar MI while the tidal forces effect the transverse components of the MI. The changes indicate that over the medium term, the Earth's polar MI decreases slightly, resulting the Earth's rotation rate increasing, albeit by a very small amount. An increase in the Earth's rotation rate has been observed since 2020, although it has been steadily decreasing over the last millennium. Thus the complexities of these changes are quite apparent. The centrifugal effect is associated with Earth's steady rotation rate. Physically, the rotation of the planet causes the equatorial bulge and consequently there is flattening at the poles.

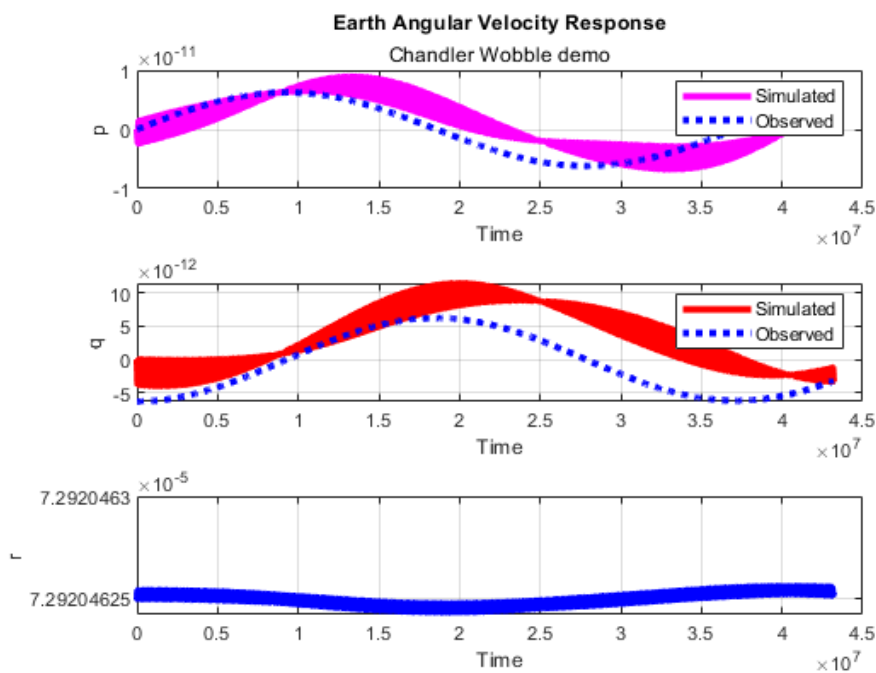

Fig. 1 Chandler Wobble: Three-axis response over 500 days

**Earth Orbit, Attitude and Rotation Rate Predictions**

The updated gravity model, could in principle, be used to predict the Earth's orbit & attitude around the Sun and hence predict changes to the Milankovitch cycles (or Earth orbit parameters) and the Earth wobble, a kind of free Eulerian nutation of the Earth's attitude caused by the redistribution of mass within the Earth or Earth's surface, using a multi-body predictor similar to the one used by JPL on their Horizons website maintained by the Solar System Dynamics Group (2025) and the method of (Patočka and Walterová 2025), (see also Yamaguchi and Furuya 2025).

The attitude dynamics response calculation method discussed in Appendix B, was used to predict the period of the Chandler wobble of the Earth which is approximately 435 days (shown in Fig. 1, along with the observed period to be about 430.3 days) and could be considered to be a relatively **short term effect**. The details of this prediction are discussed in the next sections. The observed oscillation plot is based on the period of the oscillation and the Quality factor. The increase in the sea level does not seem to significantly affect the period of Chandler wobble as the corresponding MI changes are assumed to be uniform. However, with the passage of time, this could change.

In principle, the physics of the attitude and orbital variations imply that one could predict some of the climate change parameters although not all using the approach of Kiani Shahvandi et al. (2024c) at the ETH Zürich as it is not entirely physics based. However, it is not quite clear as yet, what data inputs are needed for this prediction and how this data could be assembled from the primary sources.

**Explaining the discrepancy in the Chandler Period**

Generally if, the Earth were perfectly rigid, the frequency of its wobbling motion should be at a corresponding period is 304.5 sidereal days, i.e. 303.6 days based on the current, known, moments of inertia properties of the Earth. As a rigid Earth rotates daily, an observer on it would see that both the angular momentum and velocity vectors slowly encircle the third-axis. The three vectors stay on a same rotating plane and the angles between them remain unchanged. Due mainly to the finite elasticity of the Earth's mantle, the period of the Earth's wobble is increased. The estimate according to recent observation is about 433 or 434 days. In recognition of the one who first observed this motion, it is referred to as the Chandler wobble. Several authors have sought to explain this discrepancy but as yet there is no consensus view on its physical reasons.

The estimation of the components of the MI matrix of the Earth is discussed by Shen, Chen and Sun (2008) and by Liu, Huang and Zhang (2017). The attitude motion response is reliant on this matrix. In the moments of inertia model, from Liu, Huang and Zhang (2017), the mantle seems to be included; the core is also represented, which includes the relatively flexible fluid core as well as the rigid and solid outer core. The core may be assumed to be spherical. There are still questions as to how well it has been represented. So corrections are only necessary to the moments of inertia and not to the off diagonal products of inertia. In Eubanks (1993) the parameter $I_{zz} - I_{xx}$ is given to be 2.61×1.0e35. The value of this differential parameter is 0.325794091% of $I_{xx}$. Although it is claimed by Liu, Huang and Zhang (2017) that the accuracy or error bounds of the elements of the MI matrix is less than 0.0015%, what is important for our analysis is the error bounds for the difference in the principal moments of inertia of the Earth, which based on the data available is of the order of 0.05%. This is a relatively large uncertainty.

It may be observed that there are generally two different approaches to estimating the MI matrix of the Earth, one based on gravimetry and other based on the assumption that the geometry is determined by the hydrostatic equilibrium of the Earth matter. In fact, the gravimetry based method cannot determine all the elements of the inertia matrix and it too relies on estimates of the polar MI obtained by other methods. Thus the estimates of the moments of inertia are indeed prone to errors and it must be said, that the critical parameter in the determination of the Chandler wobble, is the difference in the moments of inertia components, $I_{zz} - I_{xx}$. According to the model proposed by Liu, Huang and Zhang (2017), $I_{zz} - I_{xx}$ is given as 2.6540×1.0e35. This value is 0.33% of $I_{xx}$. Similar results were obtain earlier by Chen and Shen (2010). In general, after considering several references the value of $I_{zz} - I_{xx}$ is within, (2.6540±0.005)×1.0e35. In fact it is observed that the estimated value of $I_{zz} - I_{xx}$ has increased from the time of Eubanks (1993) and this would actually make the error in the Chandler period much worse.

It is the belief of this author, that the reason for the discrepancy lies in the assumption that the geometric shape of the Earth also determines the inertia distribution of the Earth. The assumption made about the shape of the earth need not be true in a dynamic situation, particularly since the Earth is made up of layers of deforming material with its viscoelastic properties being quite varied across the depth of each and every layer. The relatively large uncertainty in the differences in the principal Moments of Inertia of the Earth is the main reason for the difficulties in predicting the Chandler wobble. There is a need to make small but significant corrections to the overall moments of inertia of the Earth. Thus the changes in the MI matrix of the Earth due to the deformation caused by centrifugal forces and tidal effects of the Sun must necessarily be included, as these cause errors in the difference in the moments of inertia components are mainly responsible for the observed discrepancy in the Chandler period. These errors cause considerable changes to the period of the Chandler wobble and other related characteristics of the response.

Considering the polar motion of Venus, Phan and Rambaux (2025) the viscoelastic deformation of Venus's body affects polar motion, due to the centrifugal forces, which cause variations in the MI matrix. They used a Love number formalism adopted earlier by Williams et al (2001) and obtained expressions for the changes in the MI matrix of a planet as a convolution of a Love number and changes in the MI matrix in the time domain. For our purposes one could restrict our corrections to the mean value of the Love number and express the correction to the MI matrix in matrix form as,

$$\Delta \mathbf{I}_c = \bar{k}_2 \frac{\omega_{ss}^2 R_e^5}{9G} \begin{bmatrix} -2m_z & 0 & 3m_x \\ 0 & -2m_z & 3m_y \\ 3m_x & 3m_y & 4m_z \end{bmatrix}. \quad (1)$$

In the above equation, $\omega_{ss}$ is the steady rotation speed of the planet about the axis of rotation, $R_e$ is the equatorial radius of the planet, $\bar{k}_2$ is the mean Love number associated with the planet due to the centrifugal potential, $G$ is the universal gravitation constant and $m_x$, $m_y$ and $m_z$, are defined from the relation,

$$\boldsymbol{\omega} = \boldsymbol{\omega}_{ss} + \Delta \boldsymbol{\omega} = \begin{bmatrix} 0 \\ 0 \\ \omega_{ss} \end{bmatrix} + \omega_{ss} \begin{bmatrix} m_x \\ m_y \\ m_z \end{bmatrix}. \quad (2)$$

Considering various variable values of the Love number $k_2$ which are time dependent, the mean value for the Earth was taken to be $\bar{k}_2 = 0.325$. The tidal correction term to the MI matrix due to the gravitation of the Sun, with the periodic or time varying terms ignored [17], takes the form,

$$\Delta \mathbf{I}_t = \bar{k}_2 \frac{n_{orbital}^2 R_e^5}{2G} \begin{bmatrix} 1 & 0 & 0 \\ 0 & 1 & 0 \\ 0 & 0 & 0 \end{bmatrix}. \quad (3)$$

In equation (3), $n_{orbital}$ is the orbital rate of the Earth around the Sun. The corrections to the MI matrix given by equations (1) and (3) were incorporated at every time step in the numerical integration of the attitude dynamics. Both $\Delta \mathbf{I}_c$ and $\Delta \mathbf{I}_t$ are added to the MI matrix $\mathbf{I}$, as the integration of the attitude dynamics is done, at every time step. The results obtained in the Chandler period were shown in Fig. 1.

**Sustained Mass & Moment of Inertia Loss due Earth's temperature increase**

It seems quite important to incorporate the sustained loss of mass and moments of inertia due to the Earth's continued increase in temperature. The mass changes are critical to link with climate and temperature related changes that result from interactions with the atmosphere and oceans. It is possible in principle, to include mass changes from an energy balance model at the surface with atmospheric forcing, to predict sustained changes in temperature. Following the JULES approach, outlined by Mackay et al. (2025), which is a workflow for simulating glacier evolution with climate forcing using the physically based energy balance schemes, the exchange of data and feedbacks between these models is feasible. The reader is referred to Best et al. (2011) and Maussion et al. (2019) for a more detailed explanation of the models.

For our purposes, the simpler model proposed by Schwartz (2007) is adopted. Earth's climate system consists of a very close radiative balance between absorbed shortwave (solar) radiation and the longwave (thermal infrared) radiation emitted at the outer layers of atmosphere. When these are not in balance, there is a net forcing, i.e. the radiation emitted at the outer atmospheric sphere or otherwise lost, is less than the energy absorbed by the atmosphere and the Earth's surface. The difference is responsible for the net forcing in the energy balance equation.

For the 20th century, in time units of years,

$$C_p \frac{d\Delta T}{dt} = -C_p \frac{\Delta T}{\tau} + C_p \frac{T_0}{\tau} = -C_p \frac{\Delta T}{\tau} + F_0. \tag{4}$$

In (4), the radiative term is initially ignored on grounds of smallness. The time constant $\tau$ is related to the equilibrium temperature input sensitivity $\lambda_s^{-1}$, and the system heat capacity $C_p$, by the relations,

$$\lambda_s^{-1} \equiv T_0/F_0, \; \tau = \lambda_s^{-1} C_p. \tag{5}$$

Based on data from Schwartz (2007) for the 20$^{th}$ century, and from (4),

$$\frac{T_0}{\tau} = \frac{F_0}{C_p} = \frac{1.9}{16.7} = 0.1138, \; \tau = 5\,yrs, \; \Delta T_{ss} = T_0 = \tau \frac{F_0}{C_p} = 5 \times 0.1138 = 0.57. \tag{6}$$

For the 21$^{st}$ century, with the radiative term still ignored,

$$\tau \frac{d\Delta T}{dt} + \Delta T = K_{ct} T_0 \text{ or } 5 \frac{d\Delta T}{dt} + \Delta T = 0.57 K_{ct}. \tag{7}$$

In the above, $K_{ct}$ reflects the gain on the climate (atmospheric) related effect or the input to the energy balance equation from the atmosphere. Some authors call it the feedback factor. In the energy balance framework a positive feedback would result from a decrease in effective emissivity of the planet with increasing (Global Mean Surface Temperature) GMST because of increased water-vapour mixing ratio in the atmosphere, and/or an increase in planetary co-albedo due to decrease in cloudiness with increasing GMST.

For the 21$^{st}$ century a conservative estimate of the peak value of $K_{ct}$ is between 10 and 20. With the radiative term now included, (per unit area!), and assuming a radiative gain of energy,

$$\tau \frac{d\Delta T}{dt} + \Delta T = K_{ct} T_0 + \frac{\tau}{C_p} \sigma \times \varepsilon_{eff} \left(T_r^4 - T_{ss}^4\right), \; T_r = \Delta T + T_{ss}, \tag{8}$$

or, in time units of years from Schwartz (2007) and factoring the last term,

$$5 \frac{d\Delta T}{dt} + \Delta T = 0.57 K_{ct} + 0.29940 \sigma \times \varepsilon_{eff} \Delta T \left(T_r + T_{ss}\right)\left(T_r^2 + T_{ss}^2\right). \tag{9}$$

For the 21$^{st}$ century, the time constant $\tau$ is assumed not to change at the same rate as $K_{ct}$, although it is known to change but by a much smaller percentage. For this reason, it was set at $\tau = 5$ years. Changing the time unit to seconds since 1 year = 31.55692608E+06 seconds $\equiv Y_s$,

$$(5 \times Y_s) \frac{d\Delta T}{dt} + \Delta T = 0.57 K_{ct} + 0.29940 \sigma \times \varepsilon_{eff} \Delta T \left(T_r + T_{ss}\right)\left(T_r^2 + T_{ss}^2\right). \tag{10}$$

In the equation (10), $\sigma$ is the Stefan Boltzmann constant, and the effective emissivity of the Earth, $\varepsilon_{eff}$ is assumed to be constant and given as 0.78. The feedback factor $K_{ct}$ is assumed to be constant over the entire century and set equal to 5. It must be said here, that the feedback factor $K_{ct}$ could possess its own dynamical characteristics which are dependent on the length of the day but these have been ignored and it has been assumed to be a constant, to maintain the simplicity of the model. Furthermore, if the temperature changes given by $\Delta T$ are too large, equation (10) which is non-linear, could become unstable.

The mass of the Earth is assumed to be reduced by a relation similar to the linear regression equation used by Marzeion et al [23] and given by a relation for the change in mass of the Earth which could be reduced in our case to,

$$M_e = M_{e0} - \mu_{tm} A_{\text{exposed}} \Delta T_{ss}. \tag{11}$$

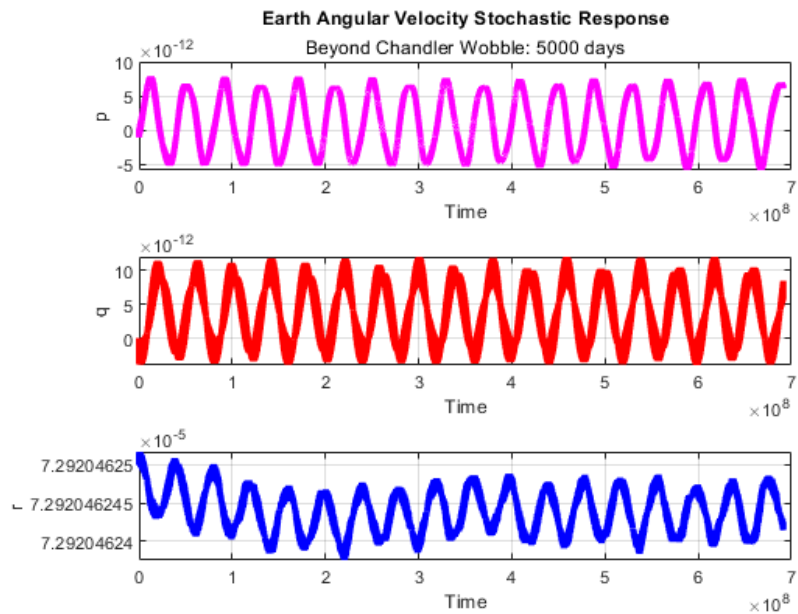

Fig. 2 Three-axis angular velocity response over the first 20 year time frame

In the above $M_e$ is the mass of the Earth at any instant of the evolution of the temperature, and $M_{e0}$ is the mass of the Earth with the absence of any temperature changes. It is assumed that the surface area of the Earth exposed to the temperature changes $A_{\text{exposed}}$, is about half the area of the oceans of the Earth, and that $\mu_{tm}$ is about 450 kg/m². It is now possible to predict the variations in the angular velocity components of the Earth over a ~20 year time frame in the 21st century. These results are shown in Figure 2.

For time frames over 20 years, results were obtained for about every 20 year time period. The general trend in the behaviour of the responses is quite similar to figure 2. The results for the time frame beyond 87.7 to 109.6 years are shown in figure 3. Moreover no instabilities were experienced over this time frame. Similar results to those shown in figures 2 and 3 were obtained, with the time constant $\tau$ set equal to 10 years. The results, both in figures 2 and 3, not only indicate a sustained decrease in the angular velocity of the Earth about the North South axis, but a ~10 year half cycle (corresponding to a ~20 year periodic response) in the same angular velocity component. Clearly this could have long term implications to climate change. So far, the cycles in the data were analysed only by visual inspection. However, a more detailed analysis of the data over the 109.6 years is currently underway using wavelet transforms and it is expected that this will provide a more detailed breakdown of the spectrum.

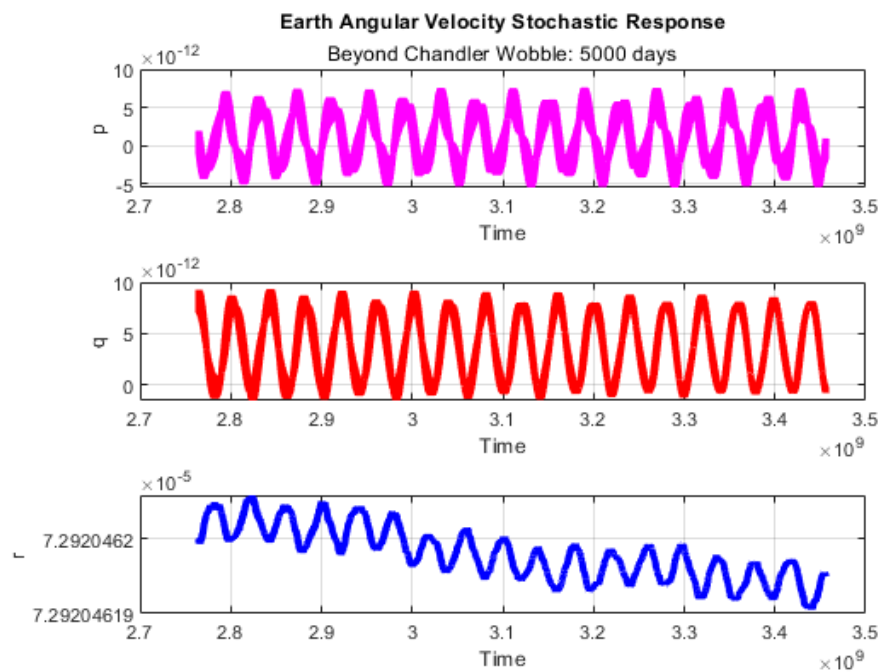

Fig. 3 Three-axis angular velocity response over a time frame from 87.7 to 109.6 years.

**Conclusions**

With the simple model presented in this paper, it is intended to highlight some of the physical features of the attitudinal motion of the planet. The reasons for the discrepancy in the period of the Chandler wobble are discussed and explained. It is indeed possible to improve the predictions, by using much more sophisticated and detailed models. For example, it is possible to introduce two equations for modelling the energy balance, one for the energy lost and a second for the energy gained. Furthermore the dynamical variations in the feedback factor $K_{ct}$ were ignored. However, the intention here has been to identify the primary physical processes that seem to determine the attitudinal motions of the planet Earth so the impact of these variations on climate change, due to the coupling of the attitudinal motion with atmospheric and oceanic changes, could then be assessed to greater degree of confidence. Moreover it was also the intention to maintain the simplicity of the complete model. A key finding of this work, is that the prediction of the attitude response of the Earth, is that the analytical model must not only include the effects of the MI changes due to the centrifugal acceleration and tidal forces generated by a deforming Earth, but also the energy balance effects which does impact on the response over the medium and long terms.

The feasibility of making predictions of Earth's rotational dynamics using accurate real-time measurements of the Earth's gravitational field, followed by some form of advanced filtering, as well as simulations to predict the orbit and attitude of the Earth itself around the Sun, is discussed. There is also a need for high fidelity modelling for climate change predictions based on physically realistic models capable of capturing the dynamics of all the key processes associated with climate change over the short, medium and long terms, so as to be able to predict realistic changes in the climate parameters during the rest of the 21$^{st}$ century. The baseline model presented in this paper can be scaled up to include more extensive thermodynamic balance equations for the atmosphere and ocean dynamics, the time dependent dynamics of the feedback factor $K_{ct}$, as well as the dynamics of the Love numbers to update the Earth's deformation dynamics.

**Acknowledgments**

## Appendix A

**Estimating the moment of inertia of the Earth due to sea level rise**

One may re-compute or estimate the MI matrix of the Earth by making the assumption that the Earth surface is an oblate spheroid, using the equations for the surface are of an oblate spheroid and the MI of an oblate spheroid shell. The density of the shell is assumed to be that of water. As only 70.8% of the Earth' surface is covered by the oceans, only 70.8% of the mass of the shell is considered. Furthermore, during this computation the mass of the Earth is assumed to be conserved. First the MI of the earth is transformed to its principal MI. The eigenvectors are saved for later use. The principal MI are reduced by the mass fraction, which is defined as the ration of 70.8% of the mass of the oblate spheroidal shell divided by the mass of the Earth. To these reduced principal MI, the principal MI of 70.8% of the mass of the oblate spheroidal shell are added, to compute the new estimated value of the principal MI of the Earth. These are then transformed back to using the saved Eigenvectors, which are assumed to remain unchanged.

## Appendix B
### Gravity Gradient Torque acting on the Earth

The Earth's attitude dynamics satisfies Euler's equations of motion,

$$I \frac{d}{dt}\boldsymbol{\omega} + \boldsymbol{\omega} \times I\boldsymbol{\omega} = \boldsymbol{\tau}, \quad I = M_e R_{gy}^2 \tag{B1}$$

where, $I$ is the MI matrix of the Earth about the rotation axis, $M_e$ is the mass of the Earth, $R_{gy}^2$ is the matrix of the squares of the radii of gyration, $\boldsymbol{\omega}$ is the rotation rate vector and $\boldsymbol{\tau}$ is the total dynamic torque acting on the earth about the same axis. Under steady state conditions,

$$I \frac{d}{dt}\boldsymbol{\omega}_{ss} = -\boldsymbol{\omega}_{ss} \times I\boldsymbol{\omega}_{ss} + \boldsymbol{\tau}_{ss} = \mathbf{0}. \tag{B2}$$

Hence under steady state conditions,

$$\boldsymbol{\tau}_{ss} = \boldsymbol{\omega}_{ss} \times I\boldsymbol{\omega}_{ss}. \tag{B3}$$

Thus the steady state gravity gradient torque $\boldsymbol{\tau}_{ss}$, could be estimated from $\boldsymbol{\omega}_{ss}$ and $I$ (of the Earth). Due to atmospheric and other disturbances it is known that $\boldsymbol{\omega}_{ss}$ reduces, and the period of the stellar day increases by about 5 ms $(\Delta T_{ss})$ which is as low as 5.8e-06% of the stellar day period. By, linearization, it can be shown that the magnitude of the perturbation disturbance torque is approximately given by,

$$|\Delta \boldsymbol{\tau}_d| = (\Delta T_{ss}/T_{ss})\boldsymbol{\tau}_{ss}, \tag{B4}$$

where $T_{ss} = 2\pi/\omega_{ss}$, is known and is the period of a stellar day (Earth's rotation period in an Inertial frame) and which is 8.4 ms more than a sidereal day.

The perturbation attitude dynamics about the steady rotation rate is given by,

$$I \frac{d}{dt}\Delta\boldsymbol{\omega} + \boldsymbol{\omega}_{ss} \times I\Delta\boldsymbol{\omega} + \Delta\boldsymbol{\omega} \times I\boldsymbol{\omega}_{ss} = \Delta\boldsymbol{\tau}, \tag{B6}$$

where $\Delta\boldsymbol{\omega}$ is the perturbation to the rotation rate vector about the steady state rotation rate vector and $\Delta\boldsymbol{\tau}$ is the perturbation torque acting on the planet which is made up of disturbance torques and a perturbation gravity gradient torque. Thus,

$$\Delta\boldsymbol{\tau} = \Delta\boldsymbol{\tau}_d + \Delta\boldsymbol{\tau}_{gg}. \tag{B7}$$

The perturbation disturbance torque $\Delta\boldsymbol{\tau}_d$, of the required magnitude is generated using a Gaussian random number generator. It can be shown based on the use of small perturbations, that,

$$\Delta \boldsymbol{\tau}_{gg} = 3n^2 \mathbf{CIc} \equiv 3n^2 \begin{bmatrix} 0 & -c_3 & c_2 \\ c_3 & 0 & -c_1 \\ -c_2 & c_1 & 0 \end{bmatrix} \begin{bmatrix} I_{11} & I_{12} & I_{13} \\ I_{12} & I_{22} & I_{23} \\ I_{13} & I_{23} & I_{33} \end{bmatrix} \begin{bmatrix} c_1 \\ c_2 \\ c_3 \end{bmatrix}. \qquad (B8)$$

The perturbation gravity gradient torque $\Delta \boldsymbol{\tau}_{gg}$, is assumed to be zero in steady state. Alternately, any steady state bias vector is filtered out. In the above $n$ is the orbital rate of the Earth around the Sun and the vector $\mathbf{c}$ is a direction cosine unit vector pointing to the direction of the centre of the Sun. Any Euler angle sequence with non-repeated indices could be used to predict the attitude of the planet. Thus the linearized perturbation equations may then be obtained and from (B6), (B7) and (B8). The direction cosine unit vector pointing to the direction of the centre of the Sun, $\mathbf{c}$ is initially expressed in an Earth fixed Earth centred inertial frame in terms of the orbital position of the Earth relative to the Sun and the obliquity. This vector, denoted as $\mathbf{c}_0$ is then transformed to an Earth centred Earth fixed body frame, so the vector $\mathbf{c}$ is expressed as,

$$\mathbf{c} = \mathbf{T}_{BI} \mathbf{c}_0,$$

where $\mathbf{T}_{BI}$ is a coordinate transformation matrix relating the Earth centred Earth fixed *inertial* frame to an Earth centred Earth fixed *body* frame and is defined in terms of (3-2-1) Euler angle sequence by the relation,

$$\begin{bmatrix} x_b \\ y_b \\ z_b \end{bmatrix} = \begin{bmatrix} 1 & 0 & 0 \\ 0 & c\theta_1 & s\theta_1 \\ 0 & -s\theta_1 & c\theta_1 \end{bmatrix} \begin{bmatrix} c\theta_2 & 0 & -s\theta_2 \\ 0 & 1 & 0 \\ s\theta_2 & 0 & c\theta_2 \end{bmatrix} \begin{bmatrix} c\theta_3 & s\theta_3 & 0 \\ -s\theta_3 & c\theta_3 & 0 \\ 0 & 0 & 1 \end{bmatrix} \begin{bmatrix} x_i \\ y_i \\ z_i \end{bmatrix} \equiv \mathbf{T}_{BI} \begin{bmatrix} x_i \\ y_i \\ z_i \end{bmatrix}. \qquad (B10)$$

In equation (B10), $c\theta_i = \cos \theta_i$ and $s\theta_i = \sin \theta_i$.

The relationship between the Euler angle sequence (3-2-1) and the angular velocity components is,

$$\begin{bmatrix} \dot{\theta}_1 \\ \dot{\theta}_2 \\ \dot{\theta}_3 \end{bmatrix} = \begin{bmatrix} 1 & \sin \theta_1 \tan \theta_2 & \cos \theta_1 \tan \theta_2 \\ 0 & \cos \theta_1 & -\sin \theta_1 \\ 0 & \sin \theta_1 / \cos \theta_2 & \cos \theta_1 / \cos \theta_2 \end{bmatrix} \begin{bmatrix} \omega_1 \\ \omega_2 \\ \omega_3 \end{bmatrix}. \qquad (B11)$$

Any other Euler angle sequence with non-repeated indices such as the (1-2-3) sequence or the (2-3-1) sequence could be used to predict the attitude of the planet. Alternatively one could also use a quaternion formulation. In our work, the dynamics was validated with each of the three Euler sequences mentioned above although only the results associated with the (3-2-1) sequence are presented.

Restricting the forcing terms to the steady-state and perturbation components, the complete nonlinear attitude dynamics of the Earth is now expressed from equations (B1), (B3) and (B7) as,

$$\mathbf{I} \frac{d}{dt} \boldsymbol{\omega} + \boldsymbol{\omega} \times \mathbf{I} \boldsymbol{\omega} = \boldsymbol{\tau}_{ss} + \Delta \boldsymbol{\tau}_d + \Delta \boldsymbol{\tau}_{gg}, \quad \mathbf{I} = M_e R_{gy}^2 + \Delta \mathbf{I}_c + \Delta \mathbf{I}_t. \qquad (B12)$$

$\Delta \boldsymbol{\tau}_{gg}$ is given by equation (B8) and the perturbation disturbance torque $\Delta \boldsymbol{\tau}_d$, of the required magnitude is generated using a Gaussian random number generator. $\Delta \mathbf{I}_c$ and $\Delta \mathbf{I}_t$ are evaluated at each time step and given by equations (1) and (3), which include the mean value of the Love number for the Earth, $\bar{k}_2$.

Appending the thermal energy balance equation from the equation (8) and the change of the mass equation given by equation (11), to the attitude dynamics, the complete set of equations for the attitude dynamics are given by equations (B11), (B12), along with equations (8) or equation (10) and equation (11).